\documentclass[leqno,thmsa]{article}
\usepackage{amssymb,a4wide,amsmath,amsthm,epsfig,graphicx,latexsym,amsfonts,amscd}
\usepackage{hyperref}
\renewcommand{\Pr}{{\mathbb{Q}}}

\makeatletter
\begin{document}
\title{Impact of the first to default time on Bilateral CVA}
\author{Damiano Brigo\thanks{%
Corresponding author. This paper expresses the views of its authors and does
not represent the institutions where the authors are working or have worked
in the past. We thank Andrea Pallavicini and Agostino Capponi for helpful discussion on CVA.} \\
{\small Dept. of Mathematics} \\
{\small King's College, London} \and Cristin Buescu  \\
{\small Dept. of Mathematics}\\
{\small King's College, London} \and
Massimo Morini\\ {\small Banca IMI and} \\ {\small Bocconi Univ., Milan}
}
\date{First version February 1, 2011.\emph{\medskip \medskip\ }This version %
\today.}
\maketitle

\begin{abstract}
We compare two different bilateral counterparty valuation adjustment (BVA) formulas.
The first formula is an approximation and is based on subtracting the two unilateral Credit Valuation Adjustment (CVA)'s formulas
as seen from the two different parties in the transaction. This formula is only a simplified representation
of bilateral risk and ignores that upon the first default closeout proceedings are ignited.
As such, it involves double counting. We compare this formula with the fully specified bilateral risk formula, where the first to default time is taken into account.
The latter correct formula depends on default dependence between the two parties, whereas the
simplified one does not. We also analyze a candidate simplified formula in case the replacement closeout is used upon default, following ISDA's recommendations, and we find the simplified formula to be the same as in the risk free closeout case.
We analyze the error that is encountered when using the simplified formula in a couple of simple products: a zero coupon bond, where the exposure is unidirectional, and an equity forward contract where exposure can go both ways.
For the latter case we
adopt a bivariate exponential distribution due to Gumbel \cite{Gumbel} to
model the joint default risk of the two parties in the deal.
We present a number of realistic cases where the simplified formula differs
considerably from the correct one.
\end{abstract}


\medskip

\noindent\textbf{AMS Classification Codes}: 62H20, 91B70 \newline
\textbf{JEL Classification Codes}: G12, G13 \newline

\noindent \textbf{Keywords}: Credit Valuation
Adjustment, Unilateral CVA, Bilateral CVA, Simplified Bilateral CVA, Debit Valuation Adjustment, Closeout, Equity Forward Contract, Zero coupon bond, Bivariate exponential distributions, Gumbel bivariate exponential distributions.

\pagestyle{myheadings} \markboth{}{{\footnotesize  D. Brigo, C. Buescu and M. Morini.
Impact of the first to default time on Bilateral CVA}}

\section{Introduction}

Counterparty credit risk has proven to be one of the major drivers of the credit crisis. We recall as a fundamental example the several credit events which occurred in one month of 2008
(see \cite{NCJ}), involving Fannie Mae, Freddie Mac, Lehman Brothers, Washington Mutual, AIG, Merrill Lynch\footnote{Merrill-Lynch did not default, technically, but as \cite{Deventers} points out: ``Merrill Lynch would probably not have survived without borrowings from the Federal Reserve. For this reason, in Kamakura's KRIS default probability data base, Merrill Lynch is classified as a `failure' for default modeling purposes."},
Landsbanki, Glitnir and Kaupthing.
We also recall that the BIS \cite{BIS} notices that

\medskip

``Under Basel II, the risk of counterparty default and credit migration risk were addressed but mark-to-market losses due to credit valuation adjustments (CVA) were not. During the financial crisis, however, roughly two-thirds of losses attributed to counterparty credit risk were due to CVA losses and only about one-third were due to actual defaults."

\medskip

This highlights the importance of proper counterparty risk valuation.
Currently, CVA is being considered in connection with capital requirements in the
Basel III accord discussion (see \cite{BIS}).
There is also a proposal for creating financial instruments that allow to securitize the CVA for a portfolio of counterparties, where margin lending and in particular the financial products called ``margin revolvers" are discussed in connection with capital management of CVA, again in Basel III. There is a lively debate in the industry on whether an effective and safe methodology for such products is available, given the lack of consensus and problems even standard corporate CDOs on i-Traxx and CDX are facing,
see for example \cite{brigopallatorre} and \cite{BieleckiBrigo}.
However, technical literature is starting to appear on these products.
See for example \cite{albanesesec}, who adopts a global valuation model and
the cutting edge technology introduced in \cite{albanese}.
We may expect the debate to develop further.

In this paper, however, we stay with pricing of CVA without dealing with securitization. In this more traditional context, situations where only default of one of the two parties is taken into account are referred to as unilateral counterparty risk. In such cases only the default of one name impacts valuation. The resulting adjustment to the otherwise default-free price of the deal, computed by the party whose default is not considered, is termed unilateral Credit Valuation Adjustment (UCVA). Unilateral CVA has been considered for example in \cite{sorensen} and in \cite{BieleckiRutkbook}, among others. Pricing of UCVA under netting is considered for example in \cite{BrigoMas}, whereas UCVA with collateral is discussed in some stylized cases and for basic products such as forward contracts in \cite{Cher05}. Precise pricing of UCVA on several asset classes with full arbitrage free dynamic models and wrong way risk is then considered in \cite{BrigoPall07} (Interest rate swaps under netting and derivatives), \cite{BrigoChourBakkar} (Oil swaps), and \cite{Brigo08} (Credit, and CDS in particular), although these works do not account for collateralization.

Situations where only default of one of the two parties is taken into account no longer look realistic after the several credit events involving financial institutions in 2008. Inclusion of both parties defaults seems desirable, also because this is the only way both parties in the deal may agree on the counterparty risk charge.
In \cite{BrigoCapponi}, a general arbitrage-free valuation framework for bilateral counterparty default risk was formalized and studied.

The bilateral nature of counterparty risk was first considered by \cite{DuffieHuang}.  In \cite{DuffieHuang}, when counterparties have different default risk, the promised cash flows of the swap are discounted using a switching discount rate that, at any given state and time, is equal to the discount rate of the counterparty for whom the swap is currently out of the money. A general formula for bilateral counterparty risk evaluation was also given in \cite{BieleckiRutkbook}.

The ongoing financial crisis has led the Basel Committee to revisit the guidelines for OTC derivatives transactions. Beside stressing the need to correctly capture the dependence between market and credit risks, which was not adequately incorporated into the Basel II framework, other amendments have been proposed, including extending the margin period of risk for OTC derivatives, increasing the incentives to use central counterparties to clear trades, and enhancing the controls regarding the re-hypothecation and re-investment of collaterals.
In this respect, \cite{BrigoCapPalVas} analyzes collateral modeling in the context of bilateral CVA, possibly under re-hypothecation, whereas \cite{BrigoMorini} analyzes the impact of closeout conventions on the bilateral CVA calculation.

In this paper, which is a refinement of \cite{buescu}, we study yet another issue concerning bilateral counterparty risk. This concerns a simplified formula for bilateral risk that is often used in the industry, see for example \cite{Picoult}. This formula, instead of considering the full bilateral CVA framework as in \cite{BrigoCapponi} or \cite{Gregory}, is based on subtracting the unilateral CVAs from the point of view of the party who is doing the calculation.
This approach neglects to model the fact that upon the first default, closeout proceedings are started and the transaction is closed. In this sense it involves inconsistent scenarios in the two terms. Indeed, assume we are in a scenario where the counterparty defaults in one year and the investor is still solvent at that time. This implies that, in that specific scenario, the DVA term payout will be zero if we use the full bilateral formula. Instead, with the simplified formula, we will be considering a DVA payout term even in the scenario where the counterparty defaults first and the transaction will be closed with a solvent investor.

The reason why the simplified formula is popular is that it allows one to compute a bilateral CVA adjustment by resorting to unilateral ones. This way one needs not implement a bilateral CVA system, but only needs to combine the output of a unilateral CVA one. In particular, what is neglected is the default dependence between the two parties involved in the deal.

In this paper we compare the two different bilateral counterparty valuation adjustment (BCVA) formulas, the correct one and the one neglecting the first to default check and closeout. We analyze a candidate simplified formula also in case the replacement closeout is used upon default, following ISDA's recommendations (see \cite{ISDAReview}), and we find the simplified formula to be the same as in the risk free closeout case.
We analyze the error that is encountered when using the simplified formula as a replacement for the full formula in a couple of simple products: a zero coupon bond, where the exposure is unidirectional, and an equity forward contract, where exposure can go both ways. For the latter case we adopt a bivariate exponential distribution due to Gumbel (see \cite{Gumbel}) to model the joint default risk of the two parties in the deal.
We present a number of realistic cases where the simplified formula differs considerably from the correct one.

Our analysis points out that in general the simplified formula is not a good approximation for the correct formula, and that care must be taken when using it to approximate the full formula.

\section{Risk-free and substitution closeout bilateral formulas}

We consider two parties in a derivative transaction: $A$ (investor) and $B$
(counterparty). We call $\tau ^{X}$, $R^{X}$ and $L^{X}=1-R^{X}$,
respectively, the default time, the recovery and the loss given default of
party $X$, with $X\in \left\{ A,B\right\} $. The risk-free discount factor
is
\begin{equation*}
D\left( t,T\right) =e^{-\int\nolimits_{t}^{T}r\left( s\right) ds}\text{,}
\end{equation*}%
where $r\left( t\right) $ is the risk-free short-rate. We define $\Pi
_{A}\left( t,T\right) $ to be the discounted cash flows of the derivative
from $t$ to $T$ seen from the point of view of $A$, with $\Pi _{B}\left(
t,T\right) =-\Pi _{A}\left( t,T\right) $. The net present value of the
derivative at $t$ is, for party A,%
\begin{equation*}
V_{A}^{0}\left( t\right) :=\mathbb{E}_{t}\left[ \Pi _{A}\left( t,T\right) %
\right] ,
\end{equation*}%
where $\mathbb{E}_{t}$ indicates the risk-neutral expectation based on
market information up to time $t$. Notice that this, in general, includes
default monitoring, i.e. the filtration at time $t$ includes
$\sigma(\{\tau^X > u\}, u \le t)$.
We denote by $\Pr_t$ the risk neutral probability measure conditional on
the same information at time $t$.

The subscript $A$ indicates that this value is seen from the point of view
of $A$, the superscript $0$ indicates that we are considering both parties
as default-free. Obviously, $V_{B}^{0}\left( t\right) =-V_{A}^{0}\left(
t\right) $.

The early literature on counterparty risk adjustment, see for example
\cite{BrigoMas}, introduced `unilateral risk of default'. Here only the
default of counterparty $B$ is considered, while the investor $A$ is treated
as default free. Under this assumption, the adjusted net present value to $A$
is%
\begin{eqnarray}
&&V_{A}^{B}\left( t\right)
\begin{tabular}{l}
$=$%
\end{tabular}%
\mathbb{E}_{t}\left\{ 1_{\left\{ \tau ^{B}>T\right\} }\Pi _{A}\left(
t,T\right) \right\} +  \\
&&+\mathbb{E}_{t}\left\{ 1_{\left\{t < \tau ^{B}\leq T\right\} }\left[ \Pi
_{A}\left( t,\tau ^{B}\right) +D\left( t,\tau ^{B}\right) \left( R^{B}\left(
V_{A}^{0}\left( \tau ^{B}\right) \right) ^{+}-\left( -V_{A}^{0}\left( \tau
^{B}\right) \right) ^{+}\right) \right] \right\}  \notag
\end{eqnarray}%

This formula can be simplified into
\begin{eqnarray}
&&V_{A}^{B}\left( t\right)
\begin{tabular}{l}
$=$%
\end{tabular}%
V_{A}^{0}\left( t\right) -\mathbb{E}_{t}\left[ L^{B}1_{\left\{t < \tau
^{B}\leq T\right\} }D\left( t,\tau ^{B}\right) \left( V_{A}^{0}\left( \tau
^{B}\right) \right) ^{+}\right] =:V_{A}^{0}\left( t\right) -\mbox{UCVA}%
_{A}\left( t\right) .  \label{formula_unilateral_counterparty}
\end{eqnarray}%

The superscript $B$ indicates that this value allows for the risk of default
of $B$. UCVA stands for Unilateral Credit Valuation Adjustment.
Notice that we always assume both parties to be alive at $t$. The
approach is easily extended to the case when $B$ is treated as default-free,
but the default of investor $A$ is taken into account. Now the adjusted net
present value to $A$ is%
\begin{eqnarray}
&&V_{A}^{A}\left( t\right)
\begin{tabular}{l}
$=$%
\end{tabular}%
\mathbb{E}_{t}\left\{ 1_{\left\{ \tau ^{A}>T\right\} }\Pi _{A}\left(
t,T\right) \right\} +  \label{formula_unilateral_counterparty2} \\
&&+\mathbb{E}_{t}\left\{ 1_{\left\{ t <\tau ^{A}\leq T\right\} }\left[ \Pi
_{A}\left( t,\tau ^{A}\right) +D\left( t,\tau ^{A}\right) \left( \left(
V_{A}^{0}\left( \tau ^{A}\right) \right) ^{+}-R^{A}\left( -V_{A}^{0}\left(
\tau ^{A}\right) \right) ^{+}\right) \right] \right\}  \notag
\end{eqnarray}%

This formula can be simplified into

\begin{eqnarray}
&&V_{A}^{A}\left( t\right)
\begin{tabular}{l}
$=$%
\end{tabular}%
V_{A}^{0}\left( t\right) +\mathbb{E}_{t}\left[ L^{A}1_{\left\{t < \tau
^{A}\leq T\right\} }D\left( t,\tau ^{A}\right) \cdot \left( -V_{A}^{0}\left(
\tau ^{A}\right) \right) ^{+}\right] =:V_{A}^{0}\left( t\right) +\mbox{UDVA}%
_{A}\left( t\right) .  \label{formula_unilateral_counterparty22}
\end{eqnarray}%

Here UDVA stands for Unilateral Debit Valuation Adjustment. Notice that
\[ \mbox{UDVA}_A( t) = \mbox{UCVA}_B( t) \]
where UCVA$_B$ is the Unilateral Credit Valuation Adjustment computed by B when only the default of A is considered.

The extension to the most realistic case when both $A$ and $B$ can default
is less trivial. This is called 'bilateral risk of default' and it is
introduced for interest rate swaps in \cite{BieleckiRutkbook}, \cite{Picoult}
(where a simplified and approximated use of the indicators is
adopted), \cite{Gregory}, \cite{BrigoCapponi}, and \cite{BrigoPallaCR}. In these previous works the net present value
adjusted by the default probabilities of both parties is given by%
\begin{eqnarray}
&&V^{AB}_{A}\left( t\right)
\begin{tabular}{l}
$=$%
\end{tabular}%
\mathbb{E}_{t}\left\{ 1_{0}\Pi _{A}\left( t,T\right) \right\}
\label{formula_biilateral_counterparty_long} \\
&&+\mathbb{E}_{t}\left\{ 1_{A}\left[ \Pi _{A}\left( t,\tau ^{A}\right)
+D\left( t,\tau ^{A}\right) \left( \left( V_{A}^{0}\left( \tau ^{A}\right)
\right) ^{+}-R^{A}\left( -V_{A}^{0}\left( \tau ^{A}\right) \right)
^{+}\right) \right] \right\}  \notag \\
&&+\mathbb{E}_{t}\left\{ 1_{B}\left[ \Pi _{A}\left( t,\tau ^{B}\right)
+D\left( t,\tau ^{B}\right) \left( R^{B}\left( V_{A}^{0}\left( \tau
^{B}\right) \right) ^{+}-\left( -V_{A}^{0}\left( \tau ^{B}\right) \right)
^{+}\right) \right] \right\} ,  \notag
\end{eqnarray}%
where we use the following event indicators if $\tau^1 = \min(\tau^A, \tau^B)$:%
\begin{eqnarray*}
1_{0} &=&1_{\left\{ T< \tau^1 \right\} } \\
1_{A} &=& 1_{\left\{t < \tau^1 = \tau^{A}\leq T  \right\} }
\\
1_{B} &=&1_{\left\{t< \tau^1 = \tau^{B}\leq T  \right\} }\\
1_{\{\tau^1 \ge t \}} &=& 1_{0} + 1_{A} + 1_{B}  
\end{eqnarray*}%

This formula can be simplified into:
\begin{eqnarray}
&&V^{AB}_{A}(t)
\begin{tabular}{l}
$=$%
\end{tabular}%
\mathbb{E}_{t}\left\{1_{\{\tau^1 \ge t\}}\Pi _{A}\left( t,T\right) \right\}
\label{formula_biilateral_counterparty_long2} \\
&&+\mathbb{E}_{t}\{ 1_{A} L^A  D(t,\tau^A)  (- V_{A}^{0}( \tau ^{A}))^{+}\}  -\mathbb{E}_{t}\{ 1_{B} L^B D( t,\tau ^{B})(  V_{A}^{0}( \tau
^{B}))^{+} \} .  \notag
\end{eqnarray}%

Notice that
\begin{equation*}
V^{AB}_{B}\left( t\right) =-V^{AB}_{A}\left( t\right) \text{,}
\end{equation*}%
thus this formula enjoys the symmetry property that one would expect.

In the above formula and throughout the paper we are assuming that the probability of $\tau^A = \tau^B$ is zero.
This is satisfied for a large variety of joint distributions on $\tau^A,\tau^B$, with important exceptions, such as the multivariate exponential Marshall Olkin distribution. The reasons why we exclude $\tau^A = \tau^B$ is that i) in practice, it is very rare that two entities default exactly at the same instant, and ii) it is not clear what the liquidation procedures would be in such a case. Therefore, to avoid ambiguity, we assume the default times are never equal. One can still follow the other quite closely and with high probability, but they never coincide.

Going back to the above bilateral formula, \cite{BrigoMorini} argue that this formula tacitly assumes that a risk free closeout is to be considered.
This is to say that upon the first default, the default-free price of the remaining deal is considered.
This is implicit in using terms such as $V_{A}^{0}\left( \tau^{A}\right)$ and $V_{A}^{0}\left( \tau^{B}\right)$ in the above formula.
The authors argue in \cite{BrigoMorini} that ISDA documentation supports another possibility, namely that the value of the remaining deal
could be computed by taking into account the credit quality of the surviving party or, in other terms, its DVA.
This is commonly referred to as ``replacement closeout" or ``substitution closeout".

The related formula is derived as:
\begin{eqnarray}
&&\hat{V}^{AB}_{A}\left( t\right)
\begin{tabular}{l}
$=$%
\end{tabular}%
\mathbb{E}_{t}\left\{ 1_{0}\Pi _{A}\left( t,T\right) \right\}
\label{formula_bilateral right} \\
&&+\mathbb{E}_{t}\left\{ 1_{A}\left[ \Pi _{A}\left( t,\tau ^{A}\right)
+D\left( t,\tau ^{A}\right) \left( \left( V_{A}^{B}\left( \tau ^{A}\right)
\right) ^{+}-R^{A}\left( -V_{A}^{B}\left( \tau ^{A}\right) \right)
^{+}\right) \right] \right\}  \notag \\
&&+\mathbb{E}_{t}\left\{ 1_{B}\left[ \Pi _{A}\left( t,\tau ^{B}\right)
+D\left( t,\tau ^{B}\right) \left( R^{B}\left( V_{A}^{A}\left( \tau
^{B}\right) \right) ^{+}-\left( -V_{A}^{A}\left( \tau ^{B}\right) \right)
^{+}\right) \right] \right\} .  \notag
\end{eqnarray}%

Formula~(\ref{formula_bilateral right}) was expressed in
the Appendix of \cite{BrigoMorini} in an equivalent form:
\begin{eqnarray}
\hat V_{A}^{AB}(t)&=&E_t[1_{\{\tau^1 \ge t\}}\Pi_A(t,T)]\label{formula_appendix}\\
&+&E_t\left\{1_A\;D(t,\tau^A)[L^A(\mbox{UDVA}_B(\tau^A)-V_A^0(\tau^A))^+ -\mbox{UDVA}_B(\tau^A)]\right\}\nonumber\\
&+&E_t\left\{1_B\;D(t,\tau^B)[\mbox{UDVA}_A(\tau^B)-L^B(V_B^0(\tau^B)-\mbox{UDVA}_A(\tau^B))^+]\right\},\nonumber
\end{eqnarray}

In this paper we consider a simplification of both formulas, and we set ourselves at time $t=0$.

\section{Simplified bilateral CVA formulas under risk free and substitution closeout}
We now consider a simplified version of both bilateral CVA formulas.
The simplification comes from the fact that instead of using the appropriate indicators
\begin{eqnarray*}
1_{A} = 1_{\left\{ \tau^1 = \tau^{A}\leq T  \right\} }, \ \ \
1_{B} =1_{\left\{ \tau^1 = \tau^{B}\leq T  \right\} }
\end{eqnarray*}%
use is made of the simplified indicators
\begin{eqnarray*}
1_{\left\{\tau^{A}\leq T  \right\} }, \ \ \
1_{\left\{\tau^{B}\leq T  \right\} }.
\end{eqnarray*}%
In other terms, one does not check for the first to default anymore and each term is computed as if it lived in a universe where only the default of one party were considered. Because we never check who defaults first, the simplified formulas do not depend on default dependence between A and B.
We begin with the risk free closeout case.

\subsection{Simplified formula with Risk Free Closeout}

The industry has been using at times a simplified version of formula (\ref{formula_biilateral_counterparty_long}), see for example \cite{Picoult}.
The simplification comes from the above substitution of indicators.   The bilateral CVA with risk free closeout, in this case, is simply the difference of the unilateral CVA adjustments as seen by the opposing parties. This has the great practical advantage that one only needs to implement a unilateral CVA formula such as (\ref{formula_unilateral_counterparty}), and use it in both directions to get the bilateral adjustment.  The simplified (`s') formula can be written as
\begin{eqnarray}
&&V^{AB,s}_{A}(0) \label{formula_biilateral_counterparty_simplif}
\begin{tabular}{l}
$=$%
\end{tabular}%
\mathbb{E}_{0}\left\{\Pi _{A}\left( 0,T\right) \right\}
\\
&&+\mathbb{E}_{0}\{ 1_{\{\tau^A \le T\}} L^A  D(0,\tau^A)  (- V_{A}^{0}( \tau ^{A}))^{+}\}  -\mathbb{E}_{0}\{ 1_{\{\tau_B\le T\}} L^B D( 0,\tau ^{B})(  V_{A}^{0}( \tau
^{B}))^{+} \} ,  \notag\\
&&= \mathbb{E}_{0}\left\{\Pi _{A}\left( 0,T\right) \right\} + \mbox{UCVA}_B(0) - \mbox{UCVA}_A(0)  \notag
\end{eqnarray}%

The first objective of this paper is compare the correct formula (\ref{formula_biilateral_counterparty_long}) with the approximated formula
(\ref{formula_biilateral_counterparty_simplif}) to single out what is lost in using the simplified formula. We would like to check that the error is negligible in most situations, but this does not seem to be a case. A warning against using (\ref{formula_biilateral_counterparty_simplif}) is then necessary
in that our later example with a forward equity contract will show that the difference can be significant with respect to the notional of the contract.

\subsection{Simplified formula with Substitution Closeout}
We suggest that even under the substitution closeout, the market may be using a simplified formula. This simplified formula, however, turns out to be equivalent to the risk-free simplified one
(\ref{formula_biilateral_counterparty_simplif}) we just derived.

Indeed, one can argue for a simplified version of formula
(\ref{formula_bilateral right}) at time $t=0$ with the same substitution of indicators.
Again, the advantage would be that the simplified formula does not
depend on default dependence between A and B. The main assumption of
the simplified formula would be that each term is computed in a universe where only one of the names can default, and this name would be the name of the first to default in the full formula.

Consider formula~(\ref{formula_appendix}).
We start with the term in the second row of the formula. We replace $1_A$ with $1_{(\tau^A\leq T)}$, and for that term we do calculations as if we were in a universe where only A can default. Since only A can default, terms such as UDVA$_B$ need be zero, since they reference the default risk of B.
A similar reasoning goes for the third row of the formula, with the difference that now the only name that can default is B, so that $1_B$ is replaced by $1_{(\tau^B\leq T)}$ and UDVA$_A$ terms are set to zero. By taking $t=0$, the simplified version with substitution
closeout turns out to be identical to
(\ref{formula_biilateral_counterparty_simplif}) :
\begin{eqnarray}
\hat V_{A}^{AB,s}(0)&=&E_0[\Pi_A(0,T)]\label{formula_simpli}\\
&+&E_0\left\{1_{(\tau^A\leq T)}\;D(0,\tau^A)[L^A(-V_A^0(\tau^A))^+]\right\}\nonumber\\
&-&E_0\left\{1_{(\tau^B\leq T)}\;D(0,\tau^B)[L^B(-V_B^0(\tau^B))^+]\right\}
=V_{A}^{AB,s}(0).\nonumber
\end{eqnarray}


\section{The difference between bilateral and simplified formulas}
In the remainder of this paper we focus on $V_A^{AB,s}$ because of
(\ref{formula_simpli}).
We derive the difference between the two formulas in general by subtracting
(\ref{formula_biilateral_counterparty_simplif}) from
(\ref{formula_biilateral_counterparty_long}). Noting that, for $t=0$:
\begin{eqnarray*}
1_A-1_{\{\tau^A<T\}}=1_{\{\tau^1=\tau^A<T\}}-1_{\{\tau^A<T\}}=
-1_{\{\tau^B<\tau^A<T\}}
\end{eqnarray*}
gives:
\begin{eqnarray}
V_A^{AB}(0)-V_A^{AB,s}(0)&=&E_0[1_{\{\tau^A<\tau^B<T\}}L^BD(0,\tau^B)
(E_{\tau^B}(\Pi(\tau^B,T)))^+]\label{diff_simpli}\\
&-&E_0[1_{\{\tau^B<\tau^A<T\}}L^A D(0,\tau^A)
(-E_{\tau^A}(\Pi(\tau^A,T)))^+].\nonumber
\end{eqnarray}
The difference is due solely to the second-to-default term.

We compute the difference (\ref{diff_simpli}) for a number of products: a
Zero Coupon Bond that has
unidirectional cash-flows, and a forward equity contract with bidirectional
cash-flows. We show that 
in the former case under unidirectional cash flows,
the risk-free simplified formula behaves better than the full bilateral
formula.
We also show that in the latter, more relevant, case (based on an equity swap) 
for $S_0>>K$ and for large $\mathbb{Q}(\tau^A<\tau^B<T)$ the difference
between the bilateral and simplified formulas becomes relevant.

\subsection{The case of a Zero Coupon Bond}
We work under deterministic interest rates. We consider $P(t,T)$ held by A (lender)
who will receive the notional 1 from B (borrower) at final maturity $T$
if there has been no default of B.
We write

\[
V^{AB}_A(0) - V^{AB,s}_A(0) = L^B P(0,T)
\; \mathbb{Q} (\tau^A < \tau^B < T).
\]

The case with substitution closeout is somehow more complicated but in the end it boils down to

\[
\hat{V}^{AB}_A(0) = P(0,T) -L^B P(0,T)\; \mathbb{Q}(\tau^B <T) = V_A^B(0).
\]
which is identical to the simplified formula (\ref{formula_simpli}). This means that, for the case of a zero coupon bond, the simplified bilateral formula coincides with the full bilateral formula with substitution closeout. Hence in the replacement closeout case for zero coupon bonds the simplified formula is as good as the full replacement closeout formula.  This is not surprising, given the unilateral direction of the cash flows in a zero coupon bond.

Then, in the case of a zero coupon bond, the study of the difference between the full correct formula with risk free closeout and the simplified formula is the same as the study of the difference between the risk free closeout and replacement closeout formulas. For this study we refer to \cite{BrigoMorini}, where one can see the dramatic differences between such formulas in different scenarios for the default dependence of the two parties.

\subsection{The case of an Equity Forward Contract}
In this case the payoff at maturity time $T$ is given by $S_T-K$,
where $S_T$ is the price of the underlying equity at time $T$ and $K$ the strike price of
the forward contract (typically $K=S_0$, `at the money', or $K = S_0 / P(t,T)$, `at the money forward'). We compute in (\ref{diff_simpli}):
\[
E_t[\Pi(t,T)]=E_t[D(t,T)(S_T-K)]=S_t-P(t,T)K.
\]
\begin{eqnarray}
D^{AB} := V^{AB}_A(0) - V^{AB,s}_A(0)=A_1-A_2,
\end{eqnarray}
where
\begin{eqnarray}
A_1=E_0\left\{1_{\{\tau^A<\tau^B<T\}}L^B
D(0,\tau^B)(S_{\tau^B}-P(\tau_B,T)K)^+\right\}\\
A_2=E_0\left\{1_{\{\tau^B<\tau^A<T\}}L^A
D(0,\tau^A)(P(\tau_A,T)K-S_{\tau^A})^+\right\}
\end{eqnarray}

We can see clearly how the difference $D^{AB}$ between the two formulas is structured. Below we present a numerical example illustrating how such difference changes and highlighting cases where the approximation is very poor and should not be used, even when ignoring wrong way risk. We can already guess that the worst cases will be the ones where the terms $A_1$ and $A_2$ do not compensate each other, leading to a large error when using the simplified formula. Such a case occurs for example when there is a high probability that $\tau^A<\tau^B$ and when the forward contract is deep in the money, since in such case $A_1$ will be large and $A_2$ will be small. Similarly, a case where $\tau^B<\tau^A$ is very likely and where the forward contract is deep out of the money will lead to a large $A_2$ and to a small $A_1$, leading again to a large difference between the two formulas.

\subsection{A numerical case study}
We consider an equity forward contract with maturity $T$.
To focus on the essential features, we assume zero interest rates so that $D(t,T)=1$ for all $t\le T$. We assume recoveries to be zero and take the equity stock $S_t$ to be independent of the defaults $\tau_A$ and $\tau_B$.

To analyze the separate impact of credit spreads and default dependence on the approximate formula, we adopt for $\tau^A,\tau^B$ a bivariate exponential distribution. As discussed in \cite{buescu}, there are several bivariate exponential distributions that could be used. One of the most utilized bivariate exponential distributions is the Marshall-Olkin bivariate distribution (see \cite{MarshallOlkin}). This distribution generalizes the lack of memory property of univariate exponential distributions.  The Marshall Olkin bivariate distribution, however, features a singular component and admits a strictly positive probability that $\tau^A = \tau^B$. Since we have excluded perfectly simultaneous defaults, we will instead resort to a bivariate exponential distribution among the three proposed by Gumbel in \cite{Gumbel}. The first bivariate distributions in \cite{Gumbel} satisfies an alternative characterization of lack of memory, also known as bivariate remaining life constancy.  However, this distribution can only describe negative dependence and in a limited range. The second bivariate exponential in \cite{Gumbel} only describes a range of dependence $[-1/4, 1/4]$ for correlation, and as such is not suited to our purposes. This is why we resort to the third bivariate exponential distribution only briefly introduced in \cite{Gumbel}. See also \cite{LuBhatta} and \cite{Kotz}.

The joint survival function for our bivariate exponential is, for positive $\lambda$'s and $\theta  \in [1, \infty)$,
\[ \mathbb{Q}(\tau^A > x_1,\tau^B > x_2) := G(x_1,x_2) = \exp( - ( (\lambda_1 x_1 )^\theta + (\lambda_2 x_2 )^\theta )^{1/\theta} ). \]

Notice that the marginal distributions are exponential random variables with mean respectively $1/\lambda_1$ and $1/\lambda_2$. We will set $\lambda_1 = \lambda^A$, a constant default intensity for the first party A, and $\lambda_2 = \lambda^B$, a constant default intensity for the second party B.

We notice that Kendall's tau for this distribution, which is a good measure of dependence (invariant for invertible increasing transformations), is
\begin{equation}
 \tau^K(G) = 1 - 1/\theta.
\end{equation}
This confirms that $\theta=1$ characterizes the independence case, whereas $\theta \rightarrow \infty$ characterizes the co-monotonic case. This bivariate exponential distribution can therefore describe the whole range of positive dependence.

We can also notice that $\lambda$'s are pure marginal parameters, whereas $\theta$ is a pure dependence parameter. This is a bivariate distribution allowing for tail dependence, and does not have a singular component, so that there is zero probability that $\tau^A = \tau^B$, consistently with our assumption above.

For the purposes of this analysis we make the unrealistic assumption that there is no credit spread volatility. This assumption is not realistic and its negative features have been highlighted for example in \cite{Brigo08}. However, we will be able to show that the simplified formula is a very poor approximation even by resorting to such a simplified model, with no credit spread volatility and no wrong-way risk. \
We will instead analyze how the difference between the two formulas is impacted by the dependence between the defaults of A and B, and by the default intensity of a single name.

We assume that equity follows a geometric Brownian motion under the risk neutral measure given by (again, with zero interest rates in the drift)
\[ d S_t = \sigma S_t dW_t \]
where the volatility $\sigma$ is a positive constant and where $W$ is a standard Brownian motion under the risk neutral measure.

We take $S_0=1$, $T=5y$, $\sigma = 0.4$, and two possible strikes $K=0.8$, and $K=1$, while for the default probabilities we take
\[  \lambda^A = 0.1, \ \lambda^B = 0.05. \]
We analyze the pattern of differences $D^{A,B}$ between the correct formula and the simplified one as the dependence between defaults of the investor $A$ and the counterparty $B$ increases. We also analyze such difference when the credit spread of the investor $A$ grows.

\begin{figure}[thb]
\includegraphics[angle=270,width=15cm,totalheight=10cm]{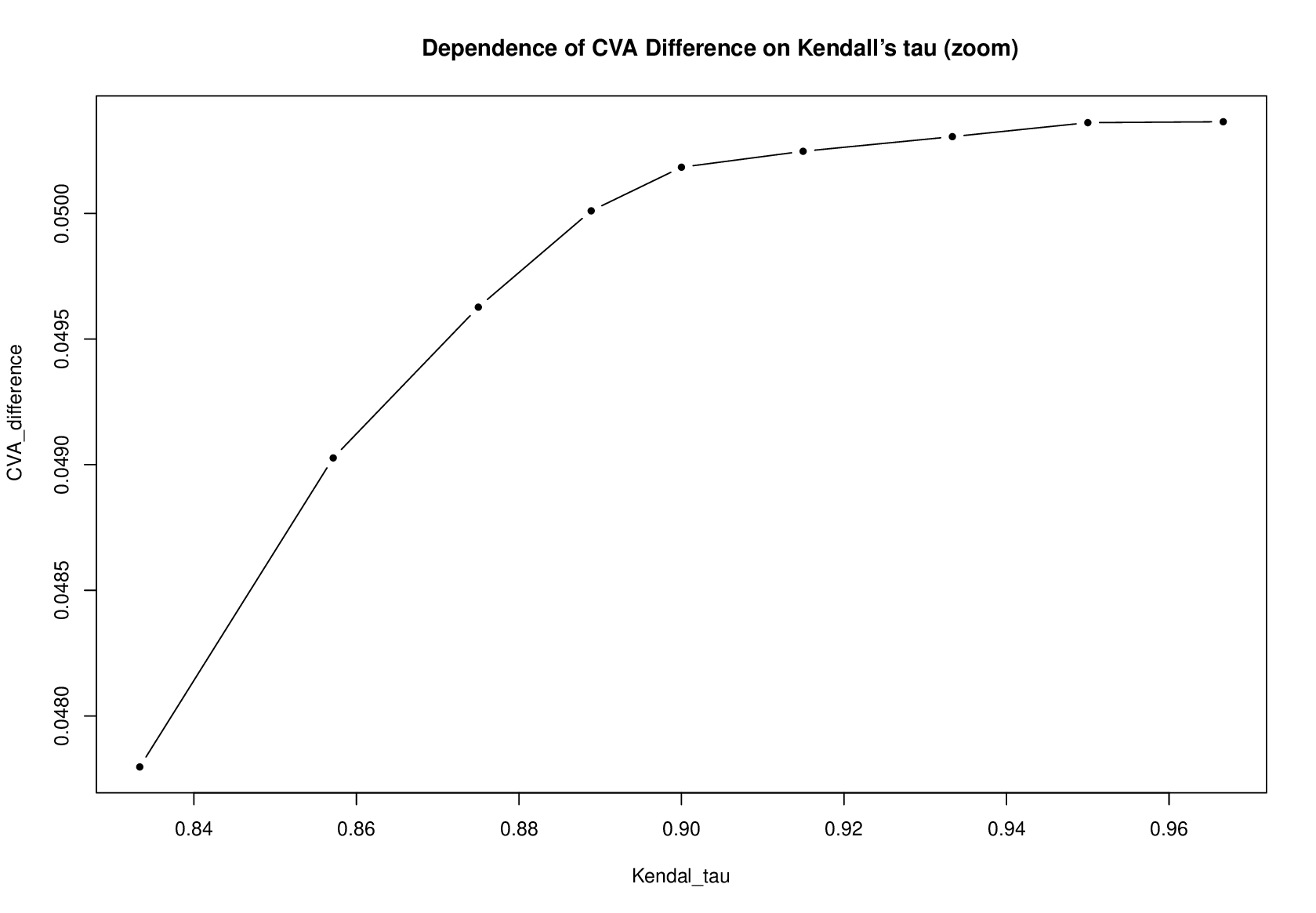}
\includegraphics[angle=270,width=15cm,totalheight=10cm]{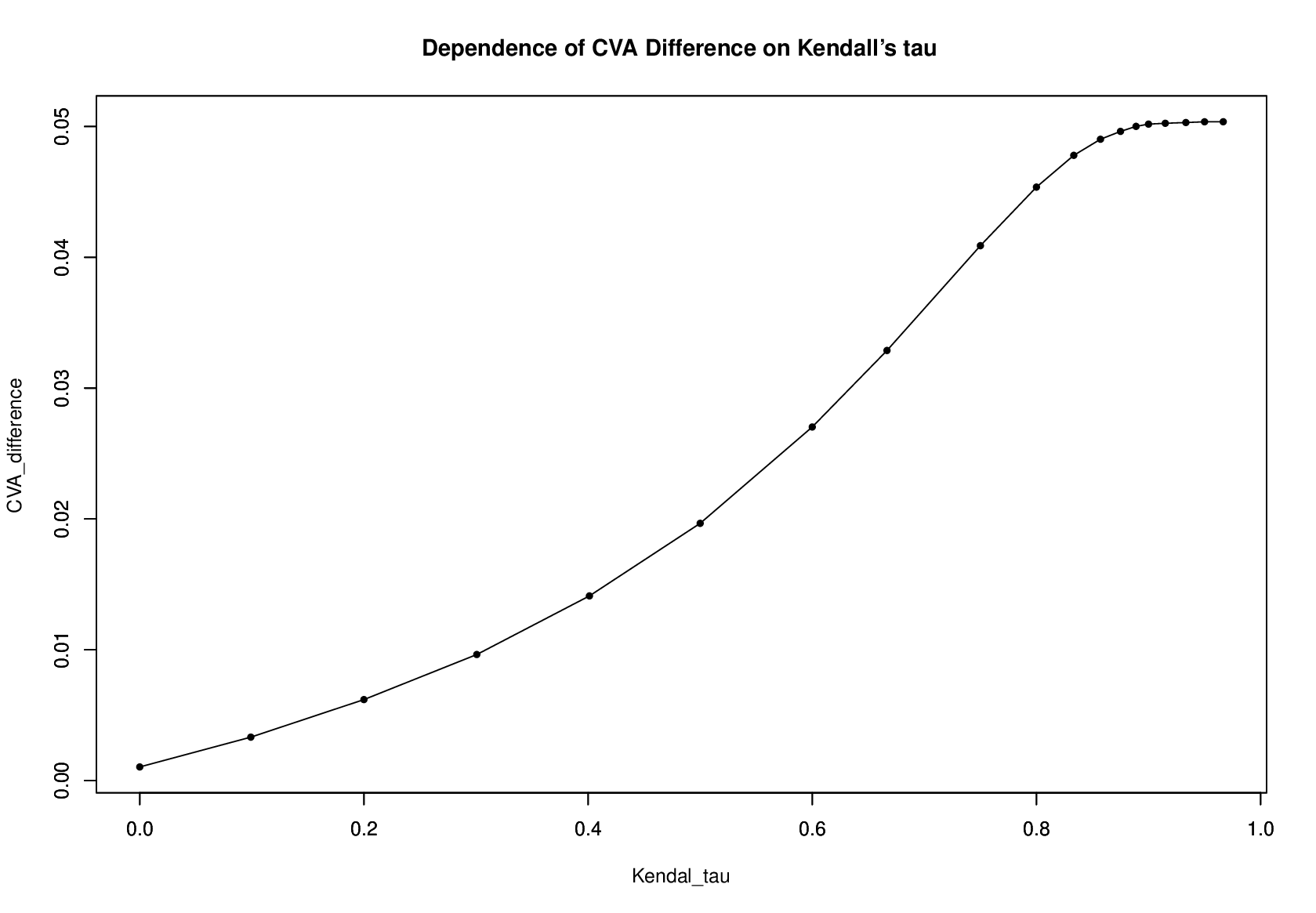}
\caption{Lower plot: $D^{AB}$ (vertical axis) plotted against Kendall's tau between
$\tau^A$ and $\tau^B$ (horizontal axis), all other quantities being equal:
$S_0=1$, $T=5y$, $\sigma = 0.4$, $K=1$, $\lambda^A = 0.1$, $\lambda^B = 0.05$.
Upper plot: zoom on $[0.83,0.97]$.}
\label{fig:dabvstau}
\end{figure}

\begin{figure}
\includegraphics[angle=270,width=15cm,totalheight=10cm]{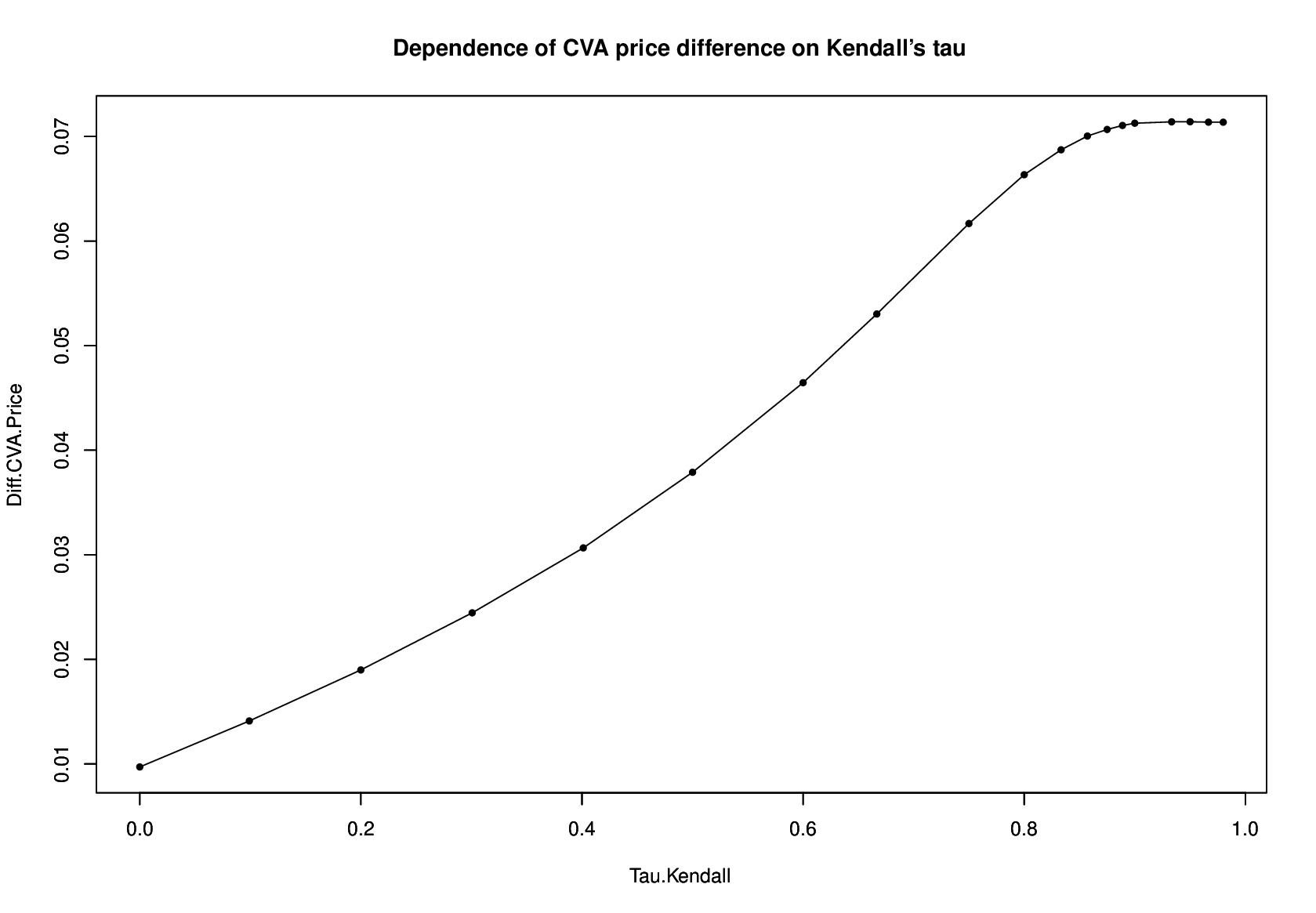}
\includegraphics[angle=270,width=15cm,totalheight=10cm]{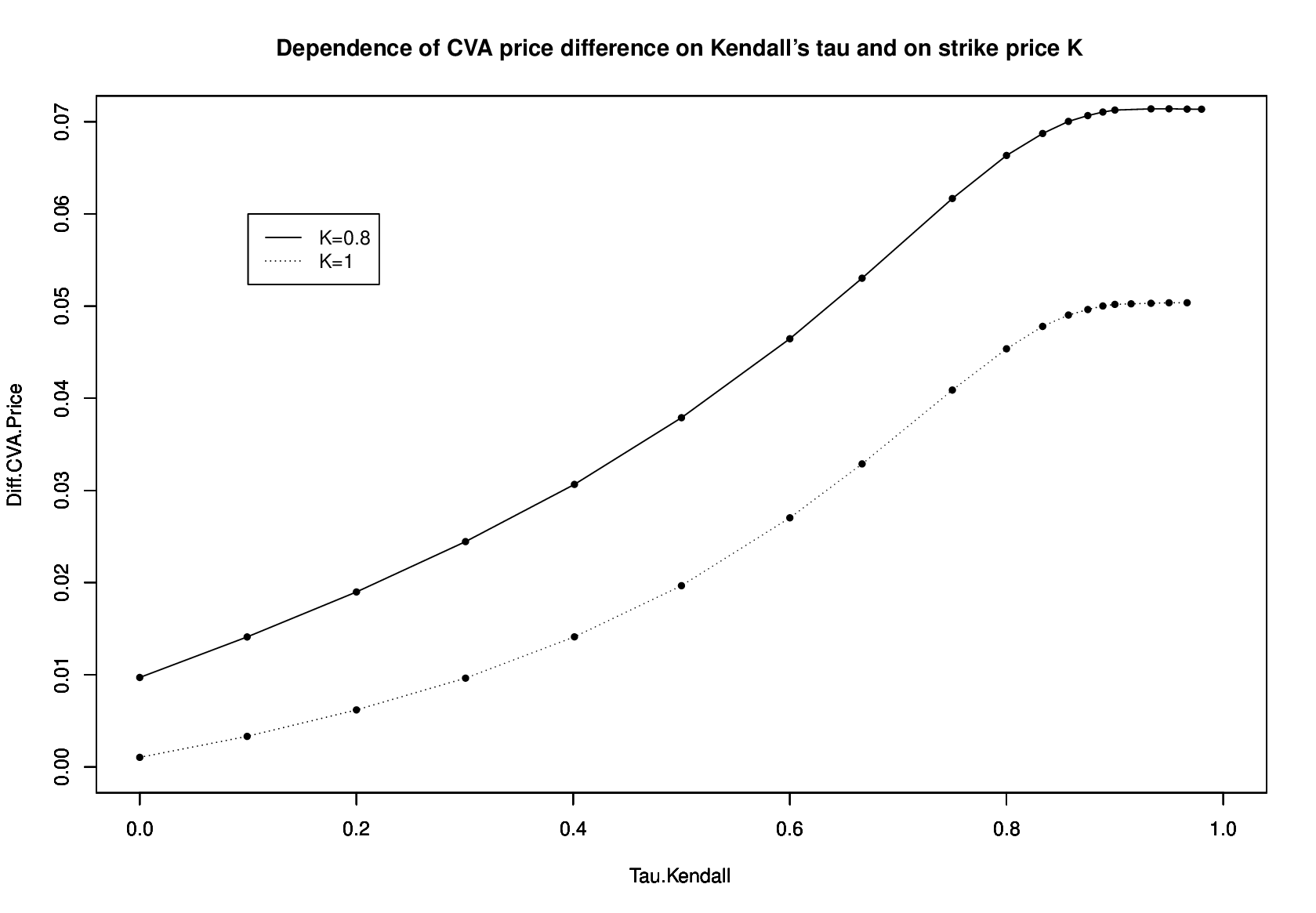}
\caption{Upper plot: $D^{AB}$ (vertical axis) plotted against Kendall's tau
between $\tau^A$ and $\tau^B$ (horizontal axis), all other quantities
being equal:
$S_0=1$, $T=5y$, $\sigma = 0.4$, $K=0.8$, $\lambda^A = 0.1$, $\lambda^B = 0.05$.
Lower plot: Comparison of $D^{AB}$ under two different strikes
$K=1$ and $K=0.8$ as a function of Kendall's tau between $\tau^A$ and $\tau^B$}
\label{fig:dabvstau2}
\end{figure}

The impact of statistical dependence between the default times, as measured by Kendall's tau $1-1/\theta$, on the difference $D^{A,B}$ between the two formulas, is illustrated in Fig. \ref{fig:dabvstau}
and \ref{fig:dabvstau2}, as first described in \cite{buescu}. We analyze forward equity contracts with two possible strikes, $K=1$ in the first case and $K=0.8$ in the second one. We notice that, depending on the forward strike, the difference of the two formulas can reach a value of $5\%$ or $7\%$ on a notional on 1, which is quite a sizeable amount. The difference appears to be monotonically increasing in the statistical dependence of the two default times for A and B as measured by Kendall's tau. We recall that such a sizeable difference has been obtained without including wrong way risk in the CVA model.

\begin{figure}
\includegraphics[angle=270,width=15cm,totalheight=10cm]{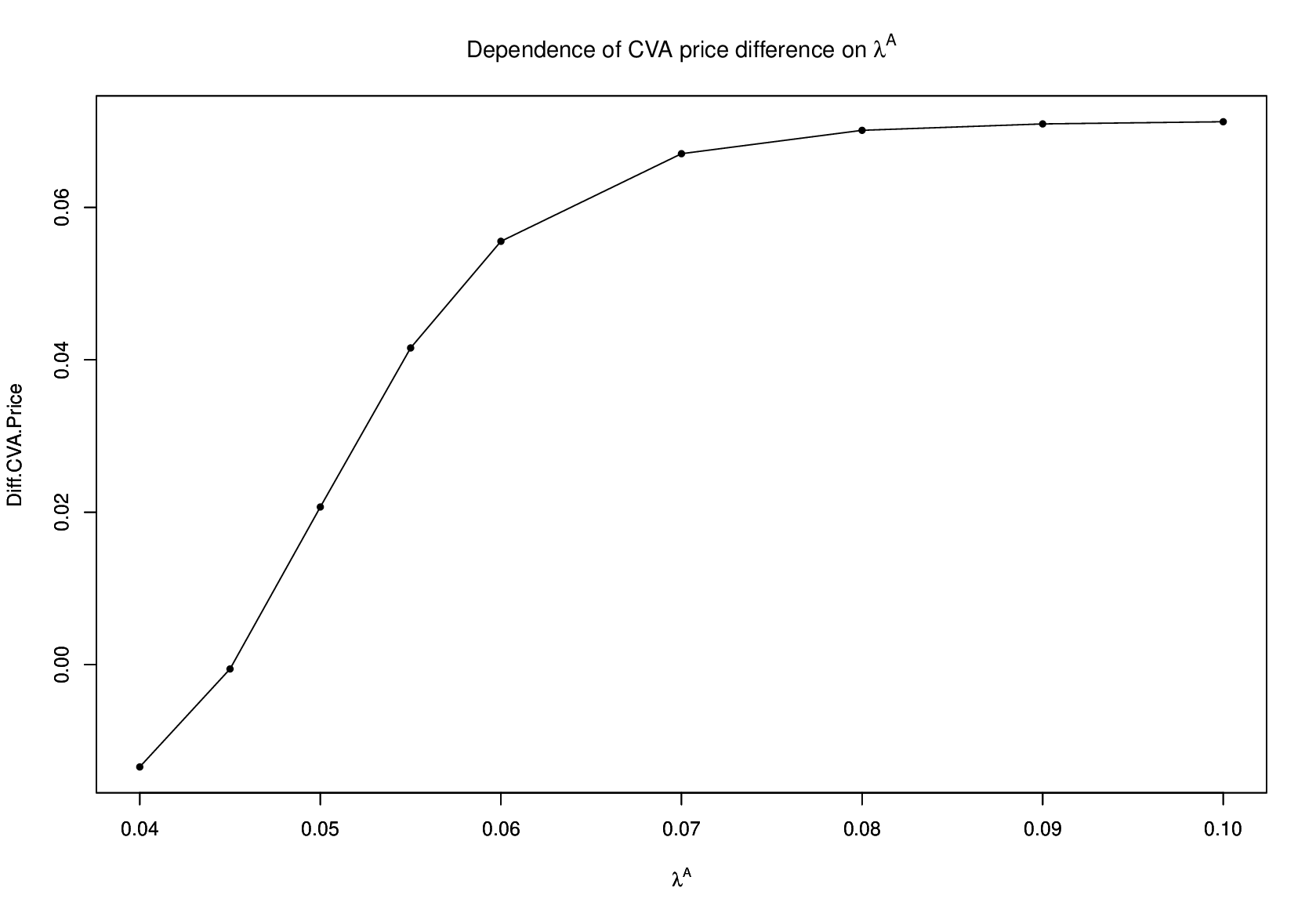}
\caption{$D^{AB}$ (vertical axis) plotted against $\lambda^A$,
default intensity of the investor $\tau^A$ (horizontal axis), all other
quantities being equal:
$S_0=1$, $T=5y$, $\sigma = 0.4$, $K=0.8$, $\lambda^B = 0.05$,
Kendall's tau = 0.9.}\label{fig:cvalambda1}
\end{figure}

We analyze the behavior of the difference as a function of $\lambda^A$ for high default dependence between A and B in Fig.~\ref{fig:cvalambda1}. We see that the difference between the two formulas increases as the intensity $\lambda^A$ increases.
However, for large values of the intensity $\lambda^A$ the increasing pattern flattens and the the difference becomes almost constant with respect to the intensity.
This is due to the fact that, for high dependence between $A$ and $B$ (Kendal's tau $= 0.9$), when the intensity of $A$ is much larger than the intensity of B and there is no credit spread volatility, then $\tau^A$ precedes $\tau^B$ in almost all scenarios. When this happens, the bilateral DVA term loses most dependence on first to default risk, since the first to default time is going to be almost always $\tau^A$. This way the bilateral DVA and the corresponding unilateral one in the simplified formula almost coincide. The bilateral CVA term is almost zero, since the event that $\tau^B$ comes before $\tau^A$ is now almost impossible. So all that is left of the difference between the two formulas is the unilateral CVA. Indeed, one can check numerically that the value to which the difference tends to grow flat asymptotically is the UCVA value, which does not depend on $\lambda^A$ or on $\theta$.

Our simulations are based on $10^{8}$ scenarios and are based on an ``R" \cite{R} source code. The maximum standard error in our simulations of the differences is $4 \times 10^{-5}$.

\section{Conclusions}
In this note we compared two different bilateral counterparty valuation adjustment formulas, the first one based on subtracting the two unilateral Credit Valuation Adjustment (CVA)'s formulas
as seen from the two different parties in the transaction, and the second one being fully inclusive of the first to default event. The first formula is only a simplified representation
of bilateral risk and ignores that upon the first default closeout proceedings are started, thus involving a degree of scenario inconsistency, while the second formula is the correct one.
The first formula is attractive because it allows for the construction of a bilateral counterparty risk pricing system based only on a unilateral one. The correct formula involves default dependence between the two parties and allows no such incremental construction, and we analyze the impact of such dependence on the difference between the two formulas. We also analyze a candidate simplified formula in case the replacement closeout is used upon default, as suggested in part of ISDA's recommendations.
We finally consider the error that is encountered when using the simplified formula in a couple of simple products: a zero coupon bond, a product with unidirectional exposure, and an equity forward contract, where exposure can go both ways. For the latter case we adopt a bivariate exponential distribution due to Gumbel (see \cite{Gumbel}) to model the joint default risk of the two parties A and B in the deal. We present a number of cases where the simplified formula differs considerably from the correct one.


\begin{thebibliography}{99}

\bibitem{albanese} Albanese, C., Bellaj, T., Gimonet, G., Pietronero, G. (2010). Coherent Global Market Simulations for Counterparty Credit Risk. Working paper available at
\emph{http://www.level3finance.com}.

\bibitem{albanesesec} Albanese, C., Pietronero, G., and S. White (2011). Optimal Funding Strategies for Counterparty Credit Risk Liabilities. Working paper available at
\emph{http://www.level3finance.com}.

\bibitem{BieleckiBrigo}
Bielecki, T., Brigo, D. and F. Patras (Editors) (2011). Credit Risk Frontiers: Sub- prime crisis, Pricing and Hedging, CVA, MBS, Ratings and Liquidity. Wiley.

\bibitem{BieleckiRutkbook}
Bielecki, T., and Rutkowski, M. (2002). Credit risk: modeling, valuation and hedging. Springer Finance, Berlin.

\bibitem{BIS}
Basel Committee on Banking Supervision, BIS (2011). Basel Committee finalises
capital treatment for bilateral counterparty credit risk. Press release
available at \emph{http://www.bis.org/press/p110601.pdf}


\bibitem{BrigoChourBakkar} Brigo, D. and I. Bakkar (2009). Accurate counterparty risk
valuation for energy-commodities swaps, {\em Energy Risk}, March 2009 issue.

\bibitem{buescu} Brigo, D. and C. Buescu (2010). Analysis of first to default risk in CVA via different multivariate exponential distributions. King's College research report, King's College, London.

\bibitem{BrigoCapponi} Brigo, D. and Capponi, A. (2008). Bilateral
counterparty risk valuation with stochastic dynamical models and application
to CDSs. Working paper available at
\emph{http://arxiv.org/abs/0812.3705}.
Short updated version in \emph{Risk Magazine}, March 2010.

\bibitem{BrigoCapPalVas}
Brigo, D., Capponi, A., Pallavicini, A., and Papatheodorou, V. (2011). Collateral Margining in Arbitrage-Free Counterparty Valuation Adjustment including Re-hypothecation and Netting. Working paper available at
\emph{arxiv.org/pdf/1101.3926}.

\bibitem{Brigo08} Brigo, D. and K. Chourdakis (2009). Counterparty Risk for
Credit Default Swaps: Impact of spread volatility and default correlation,
{\em International Journal of Theoretical and Applied Finance},  12 (07),
1007-1026.

\bibitem{BrigoMas} Brigo, D. and M. Masetti (2005). Risk Neutral Pricing of
Counterparty Risk. In
\textquotedblleft \emph{Counterparty Credit Risk
Modeling: Risk Management, Pricing and Regulation}%
\textquotedblright\
(Editor: Pykhtin, M.), Risk Books, London.

\bibitem{BrigoMorini}
Brigo, D., and Morini, M. (2010).
Dangers of Bilateral Counterparty Risk: the fundamental impact of closeout conventions.
Working paper available at
\emph{papers.ssrn.com/sol3/papers.cfm?abstract\_id=1709370}
or at
\emph{arxiv.org/pdf/1011.3355v1}.


\bibitem{BrigoPall07}
Brigo, D., and Pallavicini, A. (2007). Counterparty Risk under Correlation
between Default and Interest Rates. In
\textquotedblleft \emph{Numerical Methods for Finance}%
\textquotedblright\
(Editors: Miller, J., Edelman, D., and
Appleby, J.), Chapman Hall.

\bibitem{BrigoPallaCR} Brigo, D., Pallavicini, A., and V. Papatheodorou (2010).
Bilateral counterparty risk valuation for interest-rate products: impact of
volatilities and correlations. Working paper available at 
\emph{http://papers.ssrn.com/sol3/papers.cfm?abstract\_id=1507845}.


\bibitem{brigopallatorre} Brigo, D., Pallavicini, A., and Torresetti, R. (2010).
Credit Models and the Crisis: A journey into CDOs, Copulas, Correlations and
Dynamic Models. Wiley, Chichester.

\bibitem{Cher05}
Cherubini, U. (2005).
Counterparty Risk in Derivatives and Collateral Policies: The Replicating Portfolio Approach.
In \textquotedblleft \emph{ALM of Financial Institutions}%
\textquotedblright\  ( Editor: Tilman, L.), Institutional Investor Books.


\bibitem{DuffieHuang}
Duffie, D., and Huang, M. (1996). Swap Rates and Credit Quality, \emph{Journal of Finance} 51, 921--950.


\bibitem{Gregory} Gregory, J.K. (2009). 
Being two faced
over counterparty credit risk, 
\emph{Risk Magazine}, 22 (2),  86-90.


\bibitem{Gumbel} Gumbel, E. J. (1960). Bivariate exponential distributions, {\em Journal of the
American Statistical Association}, 55, 698-707.

\bibitem{ISDAcloseout} ISDA - International Swaps and Derivatives
Association, Inc. (2009). "ISDA close-out amount protocol", available at
\emph{www.isda.org/isdacloseoutamtprot/docs/isdacloseoutprot-text.pdf}

\bibitem{ISDAReview} ISDA - International Swaps and Derivatives Association,
Inc. (2010). Market Review of OTC Derivative Bilateral Collateralization
Practices, March 1, 2010

\bibitem{Kotz}
Kotz, S., Blakrishnan, N. and Johnson, N.L. (2000). Continuous multivariate
distributions. Wiley, New York.

\bibitem{Lindskog1} Lindskog, F., and A. McNeil (2003).
Common Poisson shock models: applications to insurance and credit risk modelling,
\emph{ASTIN Bulletin}, 33 (2), 209-238.

\bibitem{LuBhatta}
Lu, J.C. and G.K. Bhattacharyya (1991). Inference procedures for a bivariate exponential model of Gumbel based on life test of system and components, {\em J. Statist. Plann. Inference}, 27, 383–396.

\bibitem{MarshallOlkin} Marshall, A., and I. Olkin (1967). A multivariate
exponential distribution. \emph{Journal of the American Statistical
Association}, 62 (317), 30-44.

\bibitem{NCJ}
Moran, E.K. (2009). Wall Street Meets Main Street: Understanding the Financial Crisis,
N.C. Banking Inst. J., 13, 5-103, available at
 \emph{http://www.law.unc.edu/documents/journals/articles/45.pdf}.

\bibitem{moriniprampoliq} Morini, M. and A. Prampolini (2010). Risky
funding: a unified framework for counterparty and liquidity charges.
Available at \emph{http://ssrn.com/abstract=1506046}.

\bibitem{moriniRiskCopula} Morini, M. (2009). One more Model Risk when using
gaussian copula for Risk Management. Available at \emph{%
http://ssrn.com/abstract=1520670.}


\bibitem{Picoult} Picoult, E. (2005). Calculating and Hedging Exposure,
Credit Value Adjustment and Economic Capital for Counterparty Credit Risk.
In \textquotedblleft \emph{Counterparty Credit Risk Modelling}%
\textquotedblright\ (Editor: Pykhtin, M.), Risk Books.

\bibitem{R}
 R Development Core Team (2011). R: A language and environment for
  statistical computing. R Foundation for Statistical Computing,
  Vienna, Austria. ISBN 3-900051-07-0, URL http://www.R-project.org/.

\bibitem{sorensen}
Sorensen, E. H., Bollier, T.F. (1994). Pricing Swap Default Risk,
{\em Financial Analysts Journal}, 50 (3), 23-33.

\bibitem{Deventers}
Van Deventer, D. (2011). Case Studies in Liquidity Risk: Merrill Lynch. Kamakura Corporation Research. Available at
\emph{http://www.kamakuraco.com/Blog/tabid/231/EntryId/287/Case-Studies-in-Liquidity-Risk-Merrill-Lynch.aspx}
\end{thebibliography}
\end{document}